\documentclass{aa}
\usepackage{graphicx}
\usepackage[]{natbib}
\bibpunct{(}{)}{;}{a}{}{,}
\newcommand{\etal}{{\it et al.}}

\newcommand{\kms}{\mbox{km~s$^{-1}$}}

\def\sec{\ifmmode{^{\prime\prime}}\else{$^{\prime\prime}$}\fi}
\def\min{\ifmmode{^{\prime}}\else{$^{\prime}$}\fi}
\def\deg{\ifmmode{^\circ}\else{$^\circ$}\fi}
\def\arcsec#1.#2 {\ifmmode {#1^{\prime\prime}\hskip-0.42em.
                  \hskip0.15em#2}
         \else {$#1^{\prime\prime}\hskip-0.42em.\hskip0.15em#2$}
         \fi}
\def\arcmin#1.#2 {\ifmmode {#1^{\hskip 0.05em\prime}\hskip-0.35em.
                  \hskip0.05em#2}
         \else {$#1^{\hskip 0.05em\prime}\hskip-0.35em.\hskip0.05em#2$}
         \fi}
\def\arcdeg#1.#2 {\ifmmode {#1\deg\hskip-0.42em.
                  \hskip0.10em#2}
         \else {$#1\deg\hskip-0.42em.\hskip0.10em#2$}
         \fi}

\begin{document}

\title{Isolated galaxies: residual of primordial building blocks?}

\author{G. Galletta\inst{1}
        \and G. Rodighiero\inst{1}
            \and D. Bettoni\inst{2}
            \and M. Moles\inst{3}
            \and J. Varela\inst{2}}

\offprints{G. Galletta}

\institute{Dipartimento di Astronomia, Universit\`a di Padova, Vicolo
               dell'Osservatorio 2, I-35122, Padova\\
             \email{giuseppe.galletta@unipd.it}, \email{rodighiero@pd.astro.it}
         \and INAF - Osservatorio Astronomico di Padova, Vicolo dell'Osservatorio 5,
        I-35122, Padova\\
             \email{bettoni@pd.astro.it}, \email{jesus.varela@oapd.inaf.it}
              \and Instituto de Astrof\'{\i}sica de Andaluc\'{\i}a (C.S.I.C.)
               Apartado 3004, 18080 Granada, Spain\\
              \email{moles@iaa.es}
             }

\date{Received ; accepted  }

\titlerunning{Sizes of isolated galaxies}

\abstract{The mass assembly is believed to  be the dominant process of
  early  galaxy formation.  This   mechanism  of galaxy  building  can
  proceed either  by repeated major mergers  with other systems, or by
  means of accretion of matter from the surrounding regions. }
{In this  paper we compare the properties  of local disk galaxies that
  appear isolated, i.e. not  tidally affected by other galaxies during
  the last  few Gyr within the volume  given by $cz\le 5000$ \kms, with
  those galaxies at $z$ values from 0.25 to 5.}
{Effective radii for 203 isolated galaxies and  1645 galaxies from the
  RC3 have   been  collected and the two    samples have been analyzed
  statistically. A similar comparison  has  been made with half  light
  radii studied at high $z$ from the literature.}
{We  found that  isolated galaxies are   in general smaller than other
  present epoch galaxies  from the RC3.  We notice the lack of systems
  larger than 7 kpc among them.  Their size distribution appears to be
  similar to that of galaxies  at $1.4\leq z \leq  2$. The models of
  the   merging history also indicate   that the isolated galaxies did
  stop  their merging process     at  about that  redshift,   evolving
  passively since then. The   galaxy  density seems to have   remained
  unchanged since that epoch}
{Isolated  galaxies appears  to be the   end  products of the  merging
  process as proposed in the hierarchical accretion scenario at around
  $z=1.4$. For this   class of galaxies  this  was the last  significant
  merging event in their lives and  have evolved passively since then. 
  This is confirmed by the analytical estimate of the merging fraction
  with $z$ and  by  the comparison with sizes   of distant galaxies.   }


\maketitle

\begin{figure*}
\resizebox{15cm}{!}{\includegraphics[angle=0]{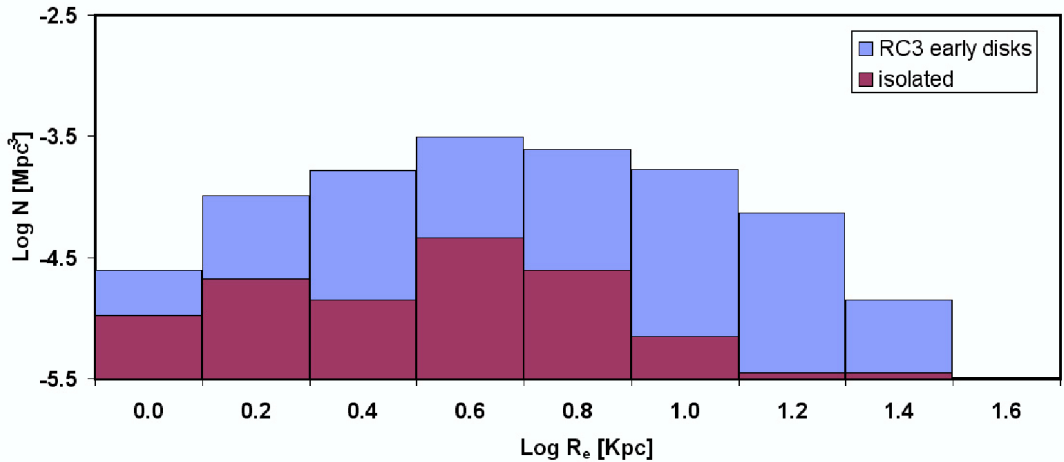}} 
\caption{Comparison between the histogram of isolated galaxies with effective radius $>$0.5 
  kpc  and the galaxies with  the same morphological type contained in
  RC3.}
\label{isto}
\end{figure*}

\section{Introduction}

The mass assembly   is believed to  be  the dominant process of  early
galaxy formation. This mechanism of galaxy building can proceed either
by repeated major mergers  with other similar  systems, or by means of
accretion of   matter  from the surrounding    regions. Some of  these
processes are  still active at  the present epoch, as  in the cases of
interacting pairs \citep{arp,am,vv,m04}, while other galaxies show the
mark of gas- or star- accretion processes happened in the past, in the
form of polar rings  or of matter rotating with  spin opposite to  the
main galaxy body    (counterrotation).  In all cases, the    accretion
process increases the final galaxy mass and powers its evolution.

The  merging or accretion processes are  in  addition independent from
the original  galaxy morphological type,  so  in principle galaxies of
every  morphological type may   increase their sizes  with the  cosmic
time, from the first, smaller building blocks to the giant galaxies at
the center of higher density environments.

According to the most extended view, galaxies  form by accretion, very
probably starting from clumpy systems similar to those observed in the
Hubble Ultra   Deep Field, later evolving  to  small exponential disks
that grow from the inside out. They start with a  dense halo and dense
disk with spiral  arms.  The exponential  structure of  these  earlier
galaxies is  predicted by  model simulations, where  different feeding
mechanisms are considered  \citep{abadi,  governato, robertson}.  Once
started,  the  small spirals should grow,   by  adding material to the
outer regions.  The  galaxy increase  his   visible size by  accreting
material and  feeding  the star formation  processes  from  the center
outwards. Often strong interactions between galaxies may destroy these
structure by  merging or cannibalising.  This picture is  supported by
the  observation of  an increase  in  size for  galaxies at  different
redshifts, as  we shall discuss  in the next sections \citep{ferguson,
  demello,  papovich}.  The building  from inside out may be supported
also by  a color gradient    detected in  galaxies at lower   redshift
\citep{papovich}.   The color evolution  according   to the process of
galaxy    formation  above   described  have    been  investigated  by
semi-analytical models   also   \citep{cattaneo}.  According to   some
authors \citep{elmegreen, noeske} the  small disk galaxies observed in
the Ultra Deep Field may be the dense,  inner regions of present-epoch
small disk galaxies.

The efficiency of the  process to build  large galaxies depends on the
initial local density, so galaxies that are initially in lower density
environments have a  lower  chance to have  large  merging events with
others galaxies or  to  accrete matter  and  should, in  principle, be
smaller and less evolved than the average.

In  a  previous  paper \citep[][  here and   after Paper  I]{jesus} we
presented a Catalogue  of disk  galaxies that  appear isolated  on the
basis of the physical criterion that they were not tidally affected by
other galaxies in their evolution during  the last few Gyr. Within the
volume given by $cz\le 5000$ \kms, 12\% of the original sample galaxies
satisfy this criterion. We found  that  they markedly differ in  size,
morphology, luminosity   and color indices  from galaxies  that are in
higher density environments.

Within  the  hierarchical  galaxy formation    frame these  properties
indicate that the building process was stopped  long time ago and they
were left  to evolve only passively from  then. In other words, if the
building of the galaxies has  proceeded by  nested merging of  smaller
building  blocks, these isolated galaxies,  selected because they were
not  involved  in significant merge events  in  the last part of their
life, are in evolutionary  stages as close as  it can be to their high
redshift counterparts.

In the present work we consider these galaxies and their properties in
the light of the recently available data on high redshifts systems, to
see  whether both  kind   of  objects  similarly reflect  the   galaxy
population at those epochs.
\section{The size of isolated galaxies}
\subsection{Comparison with normal galaxies}

In Paper I  we compared the size and  other properties of the isolated
galaxies in the Catalogue we built with that of  a sample of perturbed
galaxies,  in mild interaction with neighbors.  We found that isolated
galaxies  are smaller, bluer,  less  massive and more frequently  late
type spirals than the perturbed  galaxies.  These facts agree with the
scenario of early disks formation  described in the Introduction. Here
we  follow the analysis comparing the  size of isolated objects in our
sample  with a   larger   sample  of  stellar  systems with    similar
morphological  types extracted  from  the   RC3 \citep{rc3} and   LEDA
\citep{leda} catalogues.  We   discarded from these large samples  all
the galaxies already classified  as isolated so,  even if this  cannot
exclude   that among the galaxies present   in these huge compilations
some isolated  galaxies may exist, they  are a  fair representation of
non-isolated systems.

Isolated  galaxies in Paper  I were extracted  from the CfA2 catalogue
\citep{CfA2} with a  clear physical   criterion,  namely, to have   no
sizable companions within  a volume large  enough to ensure that  they
could not be affected in their evolution by  other systems in the last
Gyr. Other practical criteria were  1) to have the redshift available,
to establish the absolute magnitude and physical size; 2) to be within
a volume limited by $cz < 5000\,$ km/s; 3) to have morphological types
from S0 to  \mbox{Sc (-3  $\le$  t $\le$ 5);}  4)  to be  at  Galactic
latitude  $|$b$| \ge$  40\deg and  declination \mbox{$\delta\ge-2.5$.} 
All the  selected 203 galaxies have measured  diameters at  the 25 mag
arcsec$^{-2}$ isophote, D$_{25}$,    whereas only  38  have also   the
effective diameter, A$_e$, measured.

We have applied the same kind of  constraints, excluding that of being
isolated, to  the  RC3  catalogue,  in  order to   build a big  enough
comparison sample. We  found data  for  1645  systems, with 826   disk
galaxies.  Among  them only  483 (325  disks)  have measured effective
diameters. After  the conversion of the  $R_e$ radii  in kpc, we built
histograms   of size distribution taking  the  RC3 and isolated galaxy
samples binned every $\Delta$log  $R_e$=0.2 kpc and normalized  on the
volume  of  space defined  by  our   adopted selection  criteria.  The
histograms showing  the difference  in  effective radii between  local
samples (RC3 and isolated) are shown in Figure \ref{isto}, with galaxy
sizes ranging from $\sim$1 kpc to 30 kpc.

It is  clear that the two  distributions look different, since systems
with $R_e$ larger  than about  10 kpc   are very  rare among  isolated
galaxies.  This translates in  a difference in their respective median
values, that amount to 0.489 (3.08 kpc) for  the RC3 galaxies, whereas
for isolated galaxies it is only 0.369 (2.34 kpc). Both the T-test and
the F-Test give that   the difference  between  the median  values  is
statistically significant at the 95\% level.

The     clear difference in  size   distribution  between isolated and
non-isolated objects would then reflect the difference in evolutionary
histories between  both families, that,  following the standard views,
is dominated  by  merging and accretion.  If  the larger  galaxies are
formed  as  the final  step  of  a series   of collisions and  merging
processes between close systems,  this difference indicates that  this
process stopped much earlier for isolated objects than for the others.

\subsection{Merging fractions at different $z$ values}

The previous result indicates  that isolated galaxies were not altered
by any significant  merging  process for  a  large  fraction  of their
histories. To identify the  epoch  when these galaxies  were  excluded
from  the more  general  evolving  processes, we  have  considered the
models   that describe the merging  process  and its  history with the
redshift.  The merging  fraction of  galaxies  at  different epochs is
generally expressed by \citep{patton}
                       $$ f(z)= f_o \times (1+z)^m$$

\noindent
were $f_o$  and $m$ are parameters  which values change from author to
author. Thus, \citet{patton}  assumes  a constant merging fraction  of
the form $f(z)=0.011$,  while \citet{lefevre}, using more  data, finds
$f_o=0.019$ and $m=3.25$. More  recently, \citet{conselice} have found
that  both values  are  different  if massive  or small  galaxies  are
considered at different redshift ranges. They found $f_o$ ranging from
0.004 and 0.07 and $m$ ranging from 0.5 and 3.7 or higher.

Note that this   approach  assumes  a   constant local density    ($m$
constant) or describes the local  density changes assuming a different
$m$ slope  for  merging rates  with $z$.  More  accurate   models may be
considered for a fine-tuning of the resulting numbers.

Assuming the literature values, we  estimated the fraction of galaxies
that have   undergone mergers since  $z=z_m$,  following the method of
\citet{patton} by means of the formula:

$$ f(\leq z_m)_{rem}= 1- \prod_{z=0}^{z=z_m} \frac{1-f(z)}{1-0.5f(z)}$$

The percentage of surviving galaxies, that "avoided" a merger since $z_m$ is then 
$$f(\leq z_m)_{surv}=1-f(\leq z_m)_{rem}$$

\begin{table}
\caption[]{Percentage of galaxies $f(\leq z_m)_{surv}$ that survives to the merging process
at different $z$, according to two models. See the text for details. } 
\label{mergers}
      \center
            \begin{tabular}{cccccc}
      \hline
 $f_o$  &  $m$ &  $f(\leq z_m)_{rem}$ & $f(\leq z_m)_{surv}$ & $z$ & Notes\\ 
      \hline
\multicolumn{6}{l}{\citet{conselice}} \\
 0.06 & 0.5 & 88.0\% &  12.0\% & 3.8 & M$_B \le -18$\\
 0.07 & 0.4 & 88.4\% &  11.6\% & 3.7& M$_B \le -19$\\
 0.07 & 0.7 & 88.5\% &   11.5\% & 2.9 & M$_B \le -20$\\
 0.004 & 3.7 & - & - &  $>$5 & M$_B \le -21$  \\
\multicolumn{6}{l}{\citet{lefevre}} \\
 0.019 & 3.25 & 88.1\% & 11.9\% &  1.7  & - \\
        \hline

        \end{tabular}
\end{table}

We may estimate when the merging process may have  stopped for them by
calculating, with different values of $f_o$ and  $m$, at what redshift
z  the percentage of surviving galaxies  $f(\leq z_m)_{surv}$ is equal
to the observed percentage of isolated galaxies, that is $\sim$11.62\%
of the original CfA2 sample, being 203 over 1745 (Varela et al. 2004).
This is  shown in Table \ref{mergers},  assuming different  values for
$f_o$ and $m$ from  literature \citep{conselice,lefevre}. The model of
\citet{conselice} is divided in absolute magnitude bins, characterised
by different $m$ values. The values in the Table indicate that all the
galaxies have undergone several mergers since their formation, because
for  $z>$\,5   the $f$($\leq$\,$z_m$)$_{rem}$   value is greater  than
100\%.  Therefore, according to this result, our isolated galaxies may
hardly to   be residuals of the  original  blocks that  have built the
present-epoch galaxies, and none  of those blocks could  have survived
with its  pristine properties until now. However,  it appears that the
present fraction of   isolated systems is  close  to that  at redshift
between 1.7 and 3.8,  that would  imply  that the merging  process was
stopped at that epoch for such a  fraction of galaxies. This prevented
them to merge again producing larger size galaxies, as shown in Figure
\ref{isto}, and to evolve as the other, non-isolated.

\begin{figure}
\resizebox{9cm}{!}{\includegraphics[angle=0]{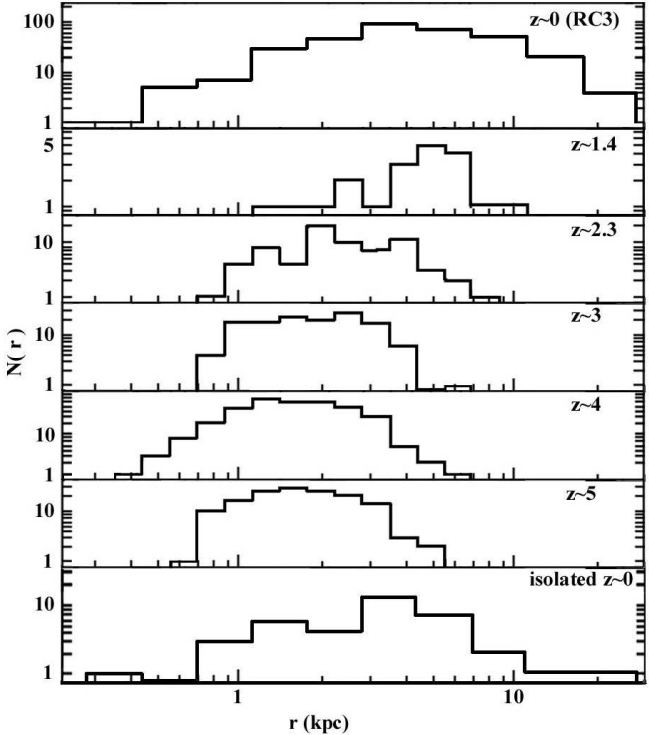}} 
\caption{The size distribution of present epoch galaxies from the RC3 (top panel) and galaxies 
  at different redshifts  from \citet{ferguson} (intermediate panels). 
  We scaled our effective radii to the  $r$ radius in arcsecond at $z=4$
  used by  these authors, to have a  direct  comparison. In the bottom
  panel we add the sizes histogram for isolated galaxies.}
\label{ferguson}
\end{figure}

\subsection{Comparison with the sizes of galaxies at higher redshifts.}

Looking back  in  time, we may  expect to  see an  epoch  in which the
galaxy size distribution  was similar to that  of present day isolated
galaxies.  This latter may be    then compared with that of   galaxies
observed at higher redshifts. \citet{ferguson, papovich, demello} have
shown that an evolution of the  physical size exists for galaxies from
$z\sim1$ to $z\sim5$. To compare our data  with these ones, we plotted
all the values in linear scale (kpc).

The  resulting histograms for local and  distant  samples are shown in
Figure \ref{ferguson}, where the sizes of the galaxies in the largest,
local existing sample are plotted together  with the sizes of galaxies
at several redshift intervals \citep{ferguson}.

Indeed,  galaxies at different   $z$ intervals are normally observed  in
different rest wavelength   intervals,  so the  comparison  is  not so
straightforward.   \citet{ferguson} selected  galaxies  from  the  HST
Advanced Camera Survey (ACS) that are observed in their UV rest-frame.
The  local  galaxies are  on  the contrary  studied  on B  images. The
\citet{ferguson} selection   may  favour   galaxies   with  large   UV
luminosities, that  are   morphologically  different from  present-day
systems and whose dimensions measured at UV band may not represent the
whole galaxy  size,   but  only   that of the   star  forming  regions
\citep{demello,  elmegreen}. However,  according  to \citet{lotz}, the
galaxy morphology does not appears to  change if galaxies are observed
between 1.5\,$\le$\,$z$\,$\le$\,4. This   means that clumpy objects  will appear
with similar morphology in this redshift intervals.

By comparing  sizes between galaxies selected  from UV data and scaled
to B-rest frame, with that of galaxies selected in  the NIR, scaled to
the same rest frame,  \cite{demello} find that  the median size of  UV
selected  galaxies appear larger   by $\sim$21\% than  that of systems
selected  in the   NIR. This may    result from selecting  in FUV  the
brightest and bigger starburst  galaxies, missing the smaller and less
active objects. However,  this is not always  true for all the samples
in the literature. Thus, the size  distribution plotted in Figure 4 of
\citet{papovich} around $z\sim 2.3$ (B-rest frame) has a  peak at 2 kpc
as that of \citet{ferguson}, but it is narrower.

Another source of error in comparing effective radii available for RC3
and isolated galaxies with the  half radii adopted by \citet{ferguson}
may   derive  from the different methods    used.  Effective radii are
computed in  large  galaxies  by means  of   integration of the  total
luminosity in nested   apertures,   using then numerical  method    to
extrapolate the total magnitude and the  radius containing half of the
luminosity. On the other side, the study  of galaxies at high redshift
and small apparent sizes is often performed with different techniques.
In the  case of the galaxies  studied by \citet{ferguson}  the program
SExtractor has  been  used, with  particular  assumptions on  the  fit
parameters \citep[see][]{ferguson}. Adopting the  same values  for the
parameters as \citet{ferguson}, we have calculated the effective radii
for the 713 galaxies observed in the field of Abell 85 cluster (Varela
et al.   in     preparation)   using SExtractor  but    also     using
GASPHOT~\citep{Pignatelli2006},    program   which performs  a surface
brightness distribution and then gives  more accurate effective radii. 
In   general,   effective  radii  calculated    with  SExtractor   are
systematically smaller than the values  obtained using other packages,
being   lower than 20\%  for  galaxies  smaller than 5   arcsec in the
analyzed images(0.4 arcsec/pixels scale).

The sizes plotted by \citet{ferguson}  derive from HST Advanced Camera
for   Surveys data with   a scale of   0.05  arcsec pix$^{-1}$ and the
apparent size  of their farthest galaxies  falls in the above range of
pixel  sizes. So, we may expect  that the histograms plotted in Figure
\ref{ferguson} may be shifted toward higher sizes  by 20\% at maximum,
or from 7 kpc to 8.4 kpc  in intrinsic size.   This shift however does
not significantly change our  results  and, therefore, we can  confirm
that large systems are very scarce in the  distribution of the size of
galaxies  in the range  1.4\,$\leq$\,$z$\,$\leq$\,5. They   are, in that sense,
similar to local isolated galaxies.

When  plotted  on the same scale,   the peak of   size distribution of
distant galaxies moves  toward higher values,  while larger and larger
galaxies are built at expenses  of the smallest stellar systems,  that
are phagocytated inside bigger ones. The  comparison with RC3 confirms
this trend.  The  present-day galaxies of    RC3 have a  peak  size at
effective  radii bigger than  galaxies at  higher $z$. Isolated galaxies
have a peak size   falling close to  that  of the distant galaxies  of
\citet{ferguson} at redshifts between 1 and 2.

In  addition  to the  shift of  the   peak  toward  larger radii  with
decreasing $z$,  in Figure \ref{ferguson} it  is visible  also a lack of
systems with  effective radii larger   than 7 kpc  (corresponding to 1
arcsec at $z=4$) in  the distant samples.  This indicates  that galaxies
larger than 7  kpc (approximately the  size of a  galaxy like M33) are
not  yet formed in the  Universe or are  very rare, still at $z=1.4$. On
the contrary in RC3, at $z=0$, they represent 22.5\% of the total and in
the isolated galaxies sample are 10.5\%.   We may deduce that isolated
galaxies have not  take part to the  merging  processes happened after
$z\ge 1$.

\begin{figure*}
\begin{center}
\resizebox{12cm}{!}{\includegraphics[angle=0]{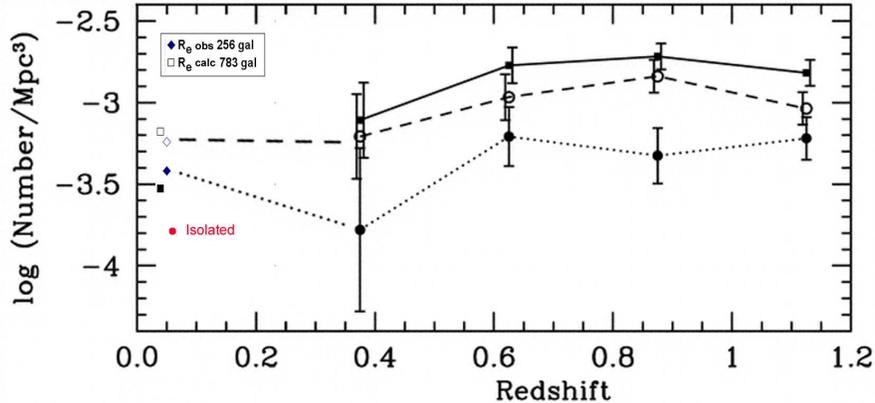}} 
\caption{Observed number densities of disk galaxies as a function of redshift. The number 
  densities of small disks ($R_e <4$\,kpc, open symbols, dashed lines)
  and large disks ($R_e  \ge 4$\,kpc,  filled symbols, dotted lines) by
  \citet{ravindra} with $z  > 0.25$ do  not show significant evolution
  in their  relative  abundance with redshift. Filled   squares, solid
  line   represent the total.   Our representative   point,  for local
  systems, has  been arbitrary plotted at  $z=0.1$, to  be visible far
  from the axis. Data from the  RC3 effective radii (256 galaxies) are
  shown as diamonds; data calculated using R$_{25}$ (783 galaxies) are
  plotted as squares.}
\label{ravindra}
\end{center}
\end{figure*}

Our data can  be  compared with  the theoretical  values of  the  mean
galaxy size with redshift  calculated assuming  different cosmological
scaling laws. These laws  describe the size  evolution in an expanding
Universe and are  related to the  assumed  cosmological geometry.  The
geometric effect    may be described  by  simple  functions (e.g. $R_e
\propto H^{-1}$  or  $R_e \propto  H^{-2/3}$) and adds  to the gradual
increase with time of  the   physical size, depicted by   hierarchical
models, due  to the accretion  of matter from  the environment or from
the merger with smaller  companions along the  galaxy lifetime. In the
literature, the  scaling laws  predict  an increase  of this parameter
with $z$ \citep{ferguson}.  This  is also  confirmed by   various recent
observations up to redshift $z\sim 5$ \citep{ferguson,cassata}.

Our result on mean  size of the whole   local disk population ($R_e  =
3.07$ kpc from RC3),  that results greater than the  mean sizes of all
the galaxies samples at  higher redshifts \citep{ferguson,cassata}, is
in   full agreement  with  the   predictions of  these  models at  low
redshift. In addition, the mean size of isolated galaxies ($R_e = 2.37$\,
kpc) is consistent with the galaxy mean size expected and observed at
$z$ between 1.5 and 3.

\subsection{Density evolution.}

We further check this  hypothesis by looking  at the density evolution
of disk galaxies.  \citet{ravindra}  have computed the observed number
density of disk  galaxies in the GOODS  CDF-S field and they found  no
significant evolution  from redshift $z=0.25$  out  to $z=1.25$.  They
divided  the sample  into small disks   ($R_e<4$\, kpc) and  large disks
($R_e \ge 4$\, kpc).  Within the uncertainties, neither sample shows a
strong evolution and remains almost constant.

The   $R_e<4$ kpc selection  criterion   for small  disks matches  the
typical size  of our isolated disks sample  (median $R_e = 2.34$ kpc). 
As we mentioned above,  along the cosmic  time smaller galaxies should
have  undergone only minor  merging   episodes,  as suggested by   the
hierarchical  growth of galaxy structures.  If  the number density  of
small disks is constant from $z=0.25$  to $z=1.25$, this might support
our suggestion on the isolated  population and on the disk  population
in general. \citet{ravindra} conclude that luminous disk galaxies were
present with roughly  the same abundances at  $z=1$ as at low redshift
($z=0.25$), and are likely to have  undergone only a modest luminosity
evolution (and then a modest number of merger events).

To better constrain the  results in the lower  redshift bin, where the
GOODS statistics is quite  poor, we adopt the  RC3 sample  to estimate
the   local   density   of    disks. Assuming   the same     value  of
\citet{ravindra}, in order to  discriminate  between small  and  large
disks,  we found  a  density  that  is still   consistent with   a non
evolutionary scenario for the density of disk  galaxies. This is shown
in  Figure \ref{ravindra},  where we  extend  the  plot of   Fig. 5 of
\citet{ravindra} to $z<0.25$.

We adopt the same  size value for  $R_e=4$ kpc to separate local small
disks (filled  lower   symbols) and local  large   disks (upper  empty
symbols).

\section{Conclusions}

{}From the  above discussion it appears  that isolated galaxies could be
important to understand the  history of the galaxy  building processes
at redshift larger than 1.4. The similarity of properties between high
redshift galaxies and the  local  isolated objects we have   selected,
indicates that,  within the frame   of hierarchical galaxy  formation,
these local isolated galaxies would have been out of the process since
that epoch.

Indeed, our isolated systems are rather small, with only a fraction of
10\% having $R_e> 7$\, kpc, at variance  with the fraction of 22.5\% that
we find in the sample of RC3 non isolated galaxies selected within the
same  volume and  with the same  specifications.  The median effective
radius amounts  to 2.3 kpc  for the  isolated  galaxies, while for RC3
this value is 3.1 kpc.

The  fraction of  isolated galaxies in   the large sample we  explored
amounts to 11.7\% (see Paper I). This figure can  be compared with the
predictions from models  of hierarchical galaxy  formation, that allow
to calculate percentage of "surviving" galaxies, that is, the fraction
of galaxies   not  affected by merging  at  a   given epoch.   Even if
different authors use   different  parameter values to describe    the
merging  evolution, we have  found  that  for  a rather large  set  of
parameters the results are similar, around 12\% at $z\sim 1.7$. This is
very close to the percentage of isolated local galaxies we found, what
let us   to conclude that  isolated galaxies  knew  their last merging
event by that epoch.

To further explore this possibility we have compared the sizes, given
by the effective radii, of the isolated galaxies with these of systems
at $z\sim1.4$ and at $z\sim2.3$. The results is that the similar sizes
are found for both families in a rather narrow z-range, close to 1.4.
Therefore, isolated galaxies with sizes larger than 7\,kpc as those in
our RC3 sample continued the building process built after $z\le1$, at
variance with our local isolated galaxies.

We also considered  the possible  density evolution  of galaxies as  a
function of their  size. We find  that neither the density of galaxies
smaller  than  $4\,$kpc, nor  that  of galaxies larger  than  $4\,$kpc have
evolved since $z =0.4$. Our RC3 data confirms  this lack of evolution to
now.

All the above results suggest that isolated galaxies were definitively
assembled at $z\sim1.4$, when the  merging process stopped due to  the
low density environment, and have evolved passively since then.

The merging history  could thus be  retraced looking at local galaxies
in gradually richer environments.


\begin{acknowledgements}
  This paper has been partially  funded by the contract  60A02-2077/05
  Fondi 2005 -  University of Padua. MM  acknowledges  a grant  by the
  Spanish   Ministerio de     Educaci\'on    y   Ciencia,    reference
  AYA2002.01241.  We would   like   to thank Dr.  Prada    for helpful
  discussion.   GR   acknowledge  Prof.   Franceschini     for  useful
  discussions.    The authors also  thanks    the  referee for  useful
  comments.
\end{acknowledgements}


%

\begin{thebibliography}{}
 
\bibitem[Abadi et al.(2003)]{abadi} Abadi, M.G., Navarro, J. F., Steinmetz M., Eke, V. R., 
2003, ApJ, 597, 21
\bibitem[Arp(1966)]{arp} Arp, H.\ 1966, Atlas of Peculiar 
Galaxies Publisher: California Institute of Technology, Pasaadena, CA, 
1966.
\bibitem[Arp \& Madore(1987)]{am} Arp, H.~C., \& Madore, 
B.~F.\ 1987, A Catalog of Southern Peculiar Galaxies and Associations 
Publisher: Cambridge University Press, 1987.
\bibitem[Cassata et al.(2005)]{cassata} Cassata, P., Cimatti, A., Franceschini, 
A., Daddi, E., Pignatelli, E., Fasano, G., Rodighiero, G., Pozzetti, L., 
Mignoli, M., Renzini, A. 2005, MNRAS, 357, 903
\bibitem[Cattaneo et al.(2006)]{cattaneo} Cattaneo, A., Dekel, A., Devriendt, J.,
Guiderdoni, B., Blaizot, J., 2006, astro-ph/0601295
\bibitem[Conselice et al.(2003)]{conselice} Conselice, C.~J., 
Bershady, M.~A., Dickinson, M., \& Papovich, C.\ 2003, AJ, 126, 1183 
\bibitem[De Mello et al.(2006)]{demello} De Mello, D. F., et al.\ 2006, AJ, 131, 216
\bibitem[de Vaucouleurs \etal (1991)]{rc3} de Vaucouleurs G., de Vaucouleurs
        A., Corwin H.G., Buta R.J., Paturel G., Fouque P., 1991, Third 
        Reference Catalogue of Bright Galaxies (RC3), Springer-Verlag: 
        New York
\bibitem[Elmegreen et al.(2005)]{elmegreen} Elmegreen, B. G. et al.\ 2005, ApJ, 634, 101
\bibitem[Ferguson et al.(2004)]{ferguson} Ferguson, H.~C., et al.\ 2004, ApJL, 600, L107 
\bibitem[Governato et al.(2004)]{governato} Governato, F., et al.\ 2004, ApJ, 607, 688
\bibitem[Huchra et al. (2000)]{CfA2} Huchra et al.\ 2000, CfA2 Catalogue of galaxy
redshift,  Center for Astrophysics.
\bibitem[Le F{\` e}vre et al.(2000)]{lefevre} Le F{\` e}vre, O., et al.\ 2000, MNRAS, 311, 565 
\bibitem[Lotz et al.(2005)]{lotz} Lotz, J.M.., et al.\ 2006, ApJ, 636, 592
\bibitem[Moles \etal (2004)]{m04} Moles, M., Bettoni, D., Fasano, G., Kj\ae rgaard, P., Varela, 
         J., 2004, A\&A, 418, 495
\bibitem[Noeske et al.(2006)]{noeske} Noeske, K. G:, et al. 2006, ApJ, 640, L143
\bibitem[Papovich et al.(2005)]{papovich} Papovich, C., et al.\ 2005, ApJ, 631, 101
\bibitem[Patton et al.(2000)]{patton} Patton, D.~R., Carlberg, 
R.~G., Marzke, R.~O., Pritchet, C.~J., da Costa, L.~N., \& Pellegrini, 
P.~S.\ 2000, ApJ, 536, 153 
\bibitem[Paturel \etal(1997)]{leda} Paturel, G., Andernach, H., Bottinelli,
      L., Di Nella, H., Durand, N., Garnier, R., Gouguenheim, L., Lanoix, P.,      
       Martinet,M.C., Petit, C., Rousseau, J., Theureau, G., Vauglin, I., 
        1997, A\&AS, 124, 109
\bibitem[Pignatelli et al.(2006)]{Pignatelli2006} Pignatelli, E.,
Fasano, G., \& Cassata, P.\ 2006, A\&A, 446, 373
\bibitem[Ravindranath et al.(2004)]{ravindra} Ravindranath, S., et al.\ 2004, ApJ, 604, L9 
\bibitem[Robertson et al.(2006)]{robertson} Robertson, B., et al.\ 2006, ApJ, in press
\bibitem[Varela et al.(2004)]{jesus} Varela, J., Moles, M., 
M{\' a}rquez, I., Galletta, G., Masegosa, J., \& Bettoni, D.\ 2004, A\&A, 
420, 873 
\bibitem[Vorontsov-Velyaminov(1959)]{vv} 
Vorontsov-Velyaminov, B.~A.\ 1959, Atlas and catalog of interacting 
galaxies (1959)
\end{thebibliography}
\end{document}